\documentclass[letterpaper,twocolumn]{esapub} 
          [1999/12/02 1.01 (PWD)]
\usepackage{times}
\usepackage{natbib}
\usepackage{graphicx}
\title{The solar tachocline and its variation (?)}
\author[1]{T. Corbard}
\author[1,2]{S.J. Jim\'enez-Reyes}
\author[1]{S. Tomczyk}
\author[1]{M. Dikpati}
\author[1]{P. Gilman}
\affil[1]{High Altitude Observatory, PO Box 3000, Boulder, CO 80307, USA}
\affil[2]{Instituto de Astrof\'\i sica de Canarias, E-38701, La Laguna, Tenerife, Spain}

\begin{document}

\keywords{Sun; Tachocline; Overshoot; Solar Cycle}

\maketitle

\begin{abstract}
The solar tachocline, located 
at the interface between the latitude-dependent 
rotation of the convection zone and the rigid radiative interior, presents 
high gradients of angular velocity which are of particular interest 
for the models of the solar dynamo and angular momentum transport. 
Furthermore, 
latitudinal and temporal variations of the tachocline parameters, if any, 
are also of particular interest in order to constrain models. 
We present a review 
of some of the theories of the tachocline and their predictions 
that may be tested 
by helioseismology. We describethe methods for inferring the tachocline parameters
from observations and the associated difficulties. A review of results 
previously obtained is given and  an analysis
of the new 6 years database of LOWL observations is presented which yields 
 no
compelling evidence of variations or general trend of the tachocline parameters
during the ascending phase of the current solar cycle (1994-2000). 
\end{abstract}

\section{Introduction}\label{sec:introduction}
The solar tachocline, so called after \citet{spiegel92},  is defined as
 the layer at the base of the convection zone (hereafter, CZ) 
where important radial  shear 
occurs. It is the transition zone where the angular velocity changes from its
latitude-dependent value in the CZ to its constant and intermediate value 
in the radiative interior.
We can emphasize three main reasons why this layer is of particular interest.

{\bf(i)}Shear turbulence and/or meridional circulation inside the tachocline may 
provide a mechanism for mixing material between the CZ and 
the radiative interior \citep[e.g.][]{brun99,schatzman00} which is needed 
to understand, for instance, the burning of lithium and the depletion of 
helium and to reach a better agreement between solar models and helioseismic
observations \citep[e.g.][]{richard96,brun99}.

{\bf(ii)}The tachocline may be the seat for angular momentum transport 
processes that could lead to the observed rigid rotation rate 
of the radiative interior. Hydrodynamical transport by unstable shear 
flow \citep[e.g.][]{chaboyer95} 
and transport by internal gravity waves in the tachocline 
\citep{kumar_talon99,kim00}
have been studied but they have not been found efficient enough 
to lead to 
an uniform internal rotation  at the present age of the Sun. This probably
requires also the presence of an internal magnetic field 
\citep{mestel_weiss,Charbonneau_mcgregor93,gough_mcintyre}. 

{\bf (iii)} Finally, the tachocline is  the best location
for an oscillatory solar dynamo which is generaly  believed to be responsible for the solar magnetic cycle for the following reasons: 
(1) Its  radial and latitudinal differential rotation  has the ability 
to produce a  
toroidal field by shearing a pre-existing poloidal field. 
(2) The $\alpha$-effect, the essential mechanism for producing poloidal 
field from toroidal field, is usually located in the CZ
but the tachocline can also produce 
a strong $\alpha$-effect  by magnetic buoyancy instability \citep{ferriz94}
  and/or by the unstable shallow-water modes \citep{dikpati_gilman_00_1}. 
(3) Because the tachocline (or  part of it)
may also be located in the slight sub-adiabatic 
overshoot layer, the toroidal fields can be stored  for an extended period 
of time, and therefore can be amplified and acted on by the $\alpha$-effect
before they escape to the surface through buoyant rise or be 
disrupted completely by convective shredding.


We will first (Sect.~\ref{sec:layers}) 
give some terminology about the different
layers and their properties at the base of the CZ 
(also to be defined). 
Then, in Sect.~\ref{sec:models}, we present 
some models of the tachocline and their predictions
that may be tested by helioseismology using methods 
presented in Sect.~\ref{sec:methods}. Previous results obtained 
are summarized in Sect~\ref{sec:eq} and we then  use the new 
LOWL data  
to investigate latitudinal 
and temporal variations of the tachocline parameters during
 the  ascending phase of the current solar cycle (Sect.~\ref{sec:variations}).

\section{Layers at the base of the convection zone: terminology, definitions and observations}\label{sec:layers}

\begin{figure}
\centering
\includegraphics[width=0.8\linewidth]{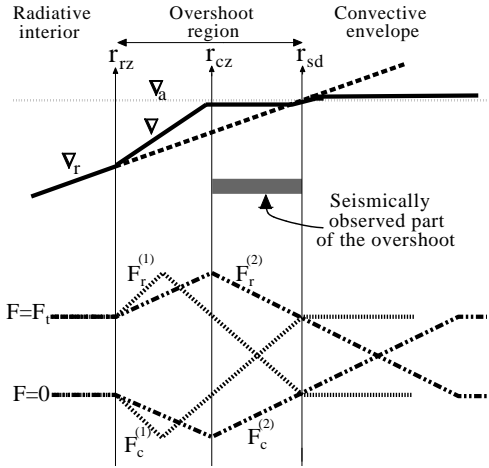}
\caption{Schematic representation of the overshoot layer 
between  the radiative zone boundary 
$r_{rz}$ and the Schwarzschild boundary $r_{sd}$ (i.e. 
where the temperature gradient 
equals both its adiabatic and radiative value $\nabla\!\!=\!\!\nabla_{a}\!\!=\!\!\nabla_{r}$). $r_{cz}$ represents the seismically observed base of the 
CZ.
 Two different models of the convective ($F_c$) and radiative ($F_r$) 
flux variations are drawn for comparison. The total flux $F_t=F_c+F_r$ 
remains constant. The model (1) of \cite{canuto97} allows  a positive 
convective flux in the slightly sub-adiabatic and 
seismically observed part of the overshoot in contradiction with 
the Schwarzschild criterion while the model (2) of \cite{zahn91} is such that
$F_c\le 0$ in all the overshoot layer.\label{fig:layers}}
\end{figure}

 As mentioned above, there is, beside the tachocline, another
layer
 defined at the interface between the super-adiabatic CZ and the radiative interior: the overshoot  layer.
Different models exist for this layer. The first particular radius, $r_{sd}$, corresponds
to the Schwarzschild boundary which defines the beginning of
the convectively unstable zone where the temperature gradient equal  its
adiabatic value $\nabla\!\!\!=\!\!\!\nabla_{ad}$.
 A second particular radius 
is where the temperature gradient is taking its radiative value ($\nabla\!=\!\nabla_{r}$) 
and  the total flux is equal to the radiative flux ($F_t\!=\!F_r$). 
This radius, $r_{rz}$, 
defines the radiative zone boundary. The region between $r_{sd}$
and $r_{rz}$ is what we here call the  overshoot region 
(see Fig.~\ref{fig:layers}). 
 The  `base of the convection zone' $r_{cz}$, measured by helioseismology,
 corresponds to the location where the 
temperature gradient changes abruptly.   
If we allow a slightly sub-adiabatic
zone below $r_{sd}$ then  helioseismology cannot
distinguish between the slightly super-adiabatic CZ and the 
slightly sub-adiabatic overshoot layer
 \citep{CD95, gilman2000} and $r_{rz}\!<\!\!r_{cz}\!<\!\!r_{sd}$. 
 Calibrations made by  \citet{basu97} lead 
to a value of $r_{cz}=0.713\pm 0.001 R_\odot$. Helioseismology can
also place an upper limit for the slightly sub-adiabatic overshoot zone 
$[r_{cz},r_{sd}]$. This upper limit is  $0.05$ pressure scale heights,
 about $2800$Km or 
$\sim 0.004R_\odot$ \citep{basu_antia97}. This is  therefore a very thin layer.
Another question concerning this layer is about the sign of the convective flux
inside. If we follow the  Schwarzschild criterion,
 we should have $F_c\!\le\! 0$ as in 
\citet{zahn91}'s model because this layer is
(slightly) sub-adiabatic,
but \citet{ferriz96} and \citet{canuto97} pointed out  
that, with a nonlocal mixing-length treatment of the CZ,
this is not necessarily
valid in the overshoot zone and we may well
have $F_c>0$ and $\nabla<\nabla_{ad}$ in spite of the Schwarzschild criterion.
 In that case the downward convective energy transport ($F_c<0$) 
occurs only in the other (lower) part of the overshoot
zone $[r_{rz},r_{cz}]$, called the overshoot layer proper \citep{canuto97, monteiro_boston98_cycle}. In the same way, 
the zone with positive convective flux is sometimes 
called CZ proper \citep[e.g.][]{ferriz96}. These differences in the
overshoot models may have some importance as it has been shown that flux tubes
with equipartition field strength ($\sim 10^4$ G) can be stored in all the overshoot
layer while the field strength of $10^5$ gauss required in order to explain
the observed active regions at the surface \citep[e.g.][]{schussler94}, 
can only be stored
where the convective flux is negative \citep{ferriz96}.

\section{Theory: some models of the tachocline and their predictions}\label{sec:models}
\subsection{Purely hydrodynamic models}
The first theory of the tachocline was developed by \citet{spiegel92}. They
mainly address the problem of the thickness of the tachocline, which we 
summarized as follows. The differential rotation $\Omega_1(\theta)$ at the top
of the radiative interior induces a latitudinal temperature gradient and
therefore a meridional circulation that would, without any stress acting on it, spread towards the interior
in a thermal time-scale leading to a non rigid rotation rate in the interior
and a very thick  tachocline for the
present Sun. We therefore need to invoke some stresses
 acting in the tachocline 
in order to prevent its progression towards the interior.
Using a 
 purely (i.e. non-magnetized) hydrodynamic model where the tachocline 
is treated as a boundary layer, \citet{spiegel92} suggest that a strong enough
 anisotropic turbulence could lead to a thin tachocline. They thus 
obtained a relation between the thickness of the tachocline $w$ and the 
horizontal turbulent viscosity coefficient $\nu_h$ assumed 
to be much higher than its vertical counterpart.
\begin{equation}\label{eq:visc}
\nu_h= 8.34\ 10^6 \left(\Omega_0\over N\right)^2\left(r_{cz}\over w\right)^4
\ \mbox{cm$^2$ s$^{-1}$}
\end{equation}

Nevertheless there remain two major questions
with this approach. First, 
 the latitudinal rotation profile inside the tachocline is likely
to be linearly stable to 2D disturbances according to Rayleigh's 
criterion  \citep{charbonneau_stability_99} and so it is probably
not the process leading to 
 the anisotropic turbulence assumed in the model.
   A solution to this problem
may be found in a recent work of \citet{dikpati_gilman_00_1}
showing that, if we allow radial deformations
of the layer and use 
the so-called shallow water model as a first approximation 
of the full 3D problem,
 then the tachocline
is found unstable in the slightly sub-adiabatic overshoot 
zone even in the purely hydrodynamic case (Fig.~{\ref{fig:stability}). 
\begin{figure}
\includegraphics[angle=90,height=0.8\linewidth]{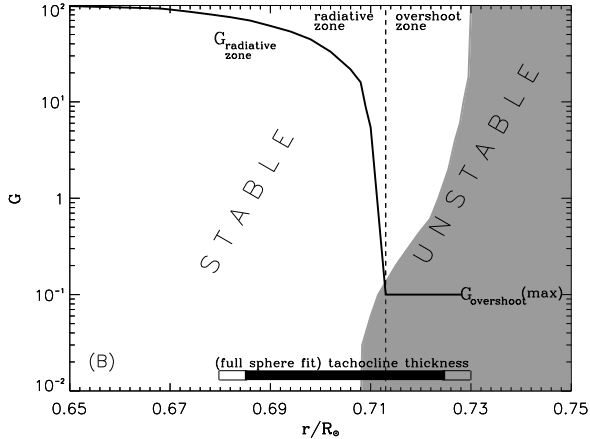}
\caption{Stability diagram from the shallow water model 
\citep{dikpati_gilman_00_1}. 
The variation of $G\simeq 10^3 |\nabla-\nabla_{ad}|$
  with radius is given
by the solid line for schematic solar model. The stability zones and 
the tachocline thickness and location  have been 
estimated from  observations 
\citep{charbonneau99,charbonneau_stability_99}. The overshoot
layer has arbitrarily been assigned a thickness of $0.015R_\odot$ and it is
slightly sub-adiabatic with $|\nabla-\nabla_{ad}|\le 10^{-4}$.
\label{fig:stability} }
\end{figure}
The second difficulty with this hydrodynamic model is that,
 even if it leads to a latitudinal independent rotation at the base of the 
tachocline, it cannot
explain the fact that there is also no significant 
radial gradient in the interior.

\subsection{Magnetized models}

Because it is very likely that the tachocline is magnetized,
 another mechanism that can lead to anisotropic turbulence 
in the tachocline is magnetic instability.
 It has effectively 
been shown  \citep{gilman_fox_1,gilman_fox_2,dikpati_gilman_99} 
that there
exists a joint instability between latitudinal differential rotation and
toroidal magnetic fields. But two different kind of magnetic fields
are usually invoked in the solar interior, namely primordial fields
and dynamo generated fields. As we shall see, 
in both cases theories relate their
properties to the shape of the tachocline.

{\bf (i)} Dynamo generated magnetic fields.
In virtually all dynamo theories (overshoot layer, interface or Babcok-Leighton
dynamos, see the review of \citet{Petrovay}), the shape 
and especially the thickness of the tachocline and the overshoot layer are
key issues  since they determine the strength of the magnetic field that can be 
stored and the process by which it is transformed. 
Using an MHD version of their shallow water model 
\citep{dikpati_gilman_00_2}
 demonstrate that the presence of a magnetic field in the tachocline 
would induce a prolateness of the tachocline otherwise oblate. This can 
also be tested by helioseismology by measuring the central position of
the radial shear at different latitudes. Such measurement
of the prolateness would allow us to estimate the strength of the magnetic
field depending on its geometry and localization inside the tachocline.
The shape of the tachocline  as obtained from this model
is shown as a function of the toroidal field strength on 
Fig~\ref{fig:prolat_th}.

\begin{figure}
\includegraphics[angle=90,height=0.8\linewidth]{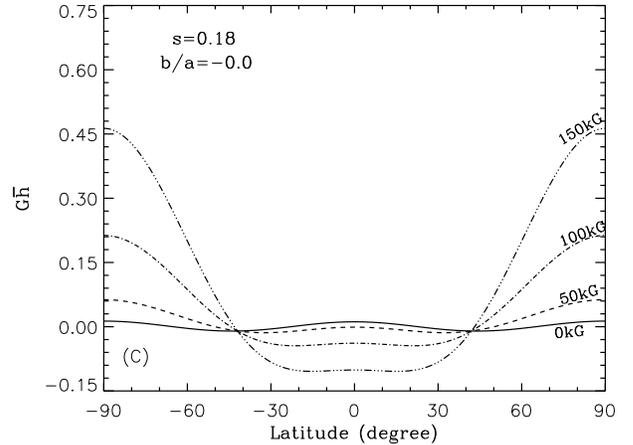}
\caption{Shape of the top surface of the tachocline with a broad nonuniform
toroidal magnetic field of different strength. G is the same quantity as in Fig.~\ref{fig:stability} and $h$ is related to the width of the tachocline by
$w\sim 0.035R_\odot(1+h)$ 
\citet{dikpati_gilman_00_2}.  \label{fig:prolat_th}}
\end{figure}

{\bf (ii)} Primordial magnetic fields.
Independently of  any magnetic field that may be 
 generated in the CZ,
\citet{gough_mcintyre} argue that we must also have a magnetic
field in the radiative interior in order to explain the uniform
(with no radial shear) rotation rate observed there. This large scale
poloidal magnetic field  would also confine 
the shear in a thin layer as a result of a balance between upward diffusion 
of the  magnetic field and downward advection by the thermally driven 
tachocline circulation.
They obtain  the following
relation between the thickness of the tachocline and the strength of the 
internal magnetic field:
\begin{equation}\label{eq:b01}
|B_0|=2.3\ 10^3 \left(\Omega_0\over N\right)^7\left(r_{cz}\over w\right)^9\ \mbox{(Gauss).}
\end{equation}
Such internal field is also required in the magnetic  models
 of \citet{rudiger} and \citet{mcgregor_charbonneau_99} where the tachocline 
is  identified with an MHD boundary layer located in the radiative interior.
 Assuming
no magnetic coupling at the core-envelope interface and that advective
effects (as modeled by \citet{spiegel92}) dominate the viscous effects 
to prevent the inward spreading of the layer, 
\citet{mcgregor_charbonneau_99} obtain:
\begin{equation}\label{eq:b02}
|B_0|\sim 3\ 10^{-8}\left(r_c\over w\right)^3 \  \mbox{(Gauss).}
\end{equation}
This work also suggests that the primordial magnetic field is  very likely
to be decoupled 
from the tachocline i.e. entirely contained within the radiative interior.
The interaction between a primordial time-independent magnetic field 
that would extend into
the CZ and the time-dependent dynamo-generated magnetic field
had nevertheless  been investigated by \citet{boruta} suggesting 
that this interaction would
lead to an inversion and an  amplification  of the primordial field by an
amount that depends again  on the thickness of the tachocline. On the other 
hand, if the dynamo  magnetic field is generated in the CZ,
it has been shown to have only a 
weak influence on the dynamics of both the tachocline and the radiative 
interior \citep{hujeirat,garaud}.

\section{HELIOSEISMOLOGY:
TOOLS AND METHODS FOR INFERRING THE
TACHOCLINE PARAMETERS}\label{sec:methods}
 One of the  observational goals
of helioseismology  is to localize precisely the tachocline with 
respect to the others described in  Sect.~\ref{sec:layers}.
 In that perspective,
we need to define clearly what we understand by tachocline parameters.

\subsection{The tachocline parameters: definition}
Following \citet{kosovichev96} we parameterize the rotation rate
 between $0.4$ and $0.8$ solar radius by an error function of the form:
\begin{equation}\label{eq:params}
\Omega(r,\theta)\!=\!\Omega_0\! +\! {{\Omega_1(\theta)-\Omega_0}\over 2} \left(\!\!1+\!erf\!\left(\!{{r -r_c(\theta)}\over 0.5\ w(\theta)}\right)\!\!\right) 
\end{equation}
where $\Omega_0$ represents the rigid
 rotation rate in the radiative interior, $\Omega_1(\theta)$ the rotation 
rate at the top of the tachocline, and  $r_c(\theta), w(\theta)$
the central position  and the width of the tachocline at the colatitude 
$\theta$. 
In order to better take into account a potential trend in the CZ
another parameter($\alpha$) 
is sometimes fitted 
by adding a term $\alpha(r-0.7)$ to Eq.~\ref{eq:params} \citep{antia98,corbard99}.
An important point to notice is
that, with this parameterization, the width $w(\theta)$ corresponds
 to a change in the rotation rate of $85\%$ of the jump 
$\Omega_1(\theta)-\Omega_0$. Other parameterizations  have been used 
\citep{antia98} but we can easily compare results using this later 
definition.
\begin{table*}
\begin{center}
\caption{Tachocline width and position as inferred from different authors, data and methods. The upper table gives values over an average in latitude  (the weighting function being $(1-\mu^2)(5\mu^2-1)$)  while the lower table is for fits at the equator. GONG month 4 starts 8/23/95. The end of GONG months 7,10 and 14 are respectively 1/13/96, 4/30/96, 9/21/96. LOWL data have been collected between 2/26/94 and 2/25/96.}

\vskip 2mm

\begin{tabular}{lllcc} 
\hline
\hline
Authors& $r_c/R_\odot$ & $w/R_\odot$ & Data& Method \\
\hline
\citet{kosovichev96}&$0.692\pm 0.005$&$0.09\pm 0.04$& BBSO 86,88-90 &FM \\
\citet{basu97}&$0.7066\pm 0.0047$&$0.0412\pm 0.0260$& BBSO 86,88-90&FM (calibration)\\
              &$0.7034\pm 0.0056$&$0.0490\pm 0.0245$& GONG 4-7 &FM (calibration)\\
              &$0.7048\pm 0.0039$&$0.0514\pm 0.0177$& GONG 4-10&FM (calibration)\\
\citet{charbonneau99}&$0.705\pm0.002$&$0.053\pm0.015$& LOWL 94-96 & Genetic FM\\
\hline
Weighted average&$0.704\pm0.002$&$0.052\pm 0.010$& &\\
\hline
\hline

\citet{charbonneau99}&$0.689\pm 0.006$&$0.01\pm 0.03$& LOWL 94-96 & OLA + Deconvolution\\
                     &$0.691\pm 0.007$&$0.07\pm 0.03$& LOWL 94-96 & RLS + Deconvolution\\
\citet{corbard98}    &$0.695\pm 0.005$&$0.05\pm 0.03$& LOWL 94-96 & RLS + Deconvolution\\
\citet{corbard99}    &$0.691\pm 0.004$&$0.01\pm 0.03$& LOWL 94-96 & Adaptive Regularization\\
\citet{antia98}      &$0.6851\pm0.0077$&$0.0230\pm0.0407$&GONG 4-14 & FM (calibration)\\
		     &$0.6843\pm0.0112$&$0.0098\pm0.0093$&GONG 4-14 & FM (simulated annealing)	\\
\hline
Weighted average& $0.691\pm0.003$& $0.034\pm0.014$& &\\
\hline
\hline
\end{tabular}
\label{tab:res}
\end{center}
\end{table*}
\subsection{The inverse problem: principle and difficulties}\label{sec:inv}
Solar oscillation modes are identified by three integers: the 
spherical harmonics degree $l$ and azimuthal order $m$, and the radial
order $n$. The frequency splitting $\delta\nu_{nlm}$
induced by the rotation $\Omega(r,\mu)$
can be calculated by:
\begin{equation}\label{eq:2D}
\delta\nu_{nlm}\!=\!\!\int_0^1\!\!\!\int_0^1\! \! K_{nlm}(r,\mu)\ \Omega(r,\mu)drd\mu
\end{equation}
where, $r$ is the fractional solar radius, $\mu=\cos(\theta)$ and 
$K_{nlm}$ are kernels calculated from a standard solar model.
Inferring the internal rotation rate from observed splittings  therefore 
requires  inverting this integral equation. In order to achieve this, 
two classes
of methods, both linear, are usually used. The first approach is a 
Least-Squares (LS) method \citep[e.g.][]{corbard97}
by which we try to find the rotation profile  that
 minimizes the sum of the square
of the differences between observed  and predicted splittings,
weighted by the observational errors.
 The second approach, called Optimally Localized 
Averages \citep[OLA, e.g.][]{pijpers92}, 
consists in  trying to reach locally, at $(r_0,\mu_0)$,
the best resolution by building the appropriate linear combination of 
observations $<\Omega(r_0,\nu_0)>=\!\!\!\sum\! c_{nlm}(r_0,\mu_0)\delta_{nlm}$. 
From Eq.~\ref{eq:2D} this is equivalent to
taking a linear combination of the kernels $K_{nlm}$. The resulting
kernel
\begin{equation}
\kappa(r_0,\mu_0,r,\mu)=\sum c_{nlm}(r_0,\mu_0)K_{nlm}(r,\mu)
\end{equation}
is called  averaging kernel because it follows from the previous
equations that the quantity $<\Omega(r_0,\mu_0)>$ is an average of
the rotation rate in a domain defined by $\kappa(r_0,\mu_0,r,\mu)$:
\begin{equation}\label{eq:convol}
 <\Omega(r_0,\mu_0)>=\int_0^1\!\!\!\int_0^1\kappa(r_0,\mu_0,r,\mu)\Omega(r,\mu)drd\mu
\end{equation}

The difficulties arise from the fact that
Eq.~\ref{eq:2D}
is an ill-posed problem with no unique solution.
In the global (LS)
 methods, we therefore need to introduce some a-priori knowledge
on the rotation profile in order to {\it regularize} the solution and avoid strong oscillations. 
Nevertheless,  this Regularized
Least Squares (RLS) approach  prevents us from recovering accurately 
sharp gradients as those expected in the tachocline because they do not
conform to the global smoothness a-priori introduced.

For local (OLA) methods, the limitation in resolution comes 
from
the error propagation. 
A quantitative measure of the resolution 
in the radial direction is obtained by fitting the averaging kernel 
by a Gaussian profile centered at $r_0$ with a standard deviation $\sigma_r$.
We then define the radial resolution by $\Delta_r\!\!=\!\!\sqrt{8 \sigma_r^2}$. 
The
reason for the choice of this definition
will become more clear in the next section but it turns out that from actual
observations the tachocline is not resolved by linear inversions when 
limiting error propagation to a reasonable amount.

Therefore  the information about the thickness of a sharp gradients
 of the rotation profile is not directly readable from 
the solutions obtained by classic inversions. 
However three different approaches have been developed in order to overcome this difficulties.

{\bf (i)} Forward analysis. The general shape of the rotation rate is assumed 
to be known and the tachocline is parameterized
 with few parameters (as the $erf$
function of Eq.~\ref{eq:params}) which are adjusted  to fit the data
by calibration \citep{basu97} or by minimization using methods such as simulated annealing \citep{antia98} or  genetic algorithm  \citep{charbonneau99}.

{\bf (ii)} Deconvolution. This method 
 uses our knowledge of the resolution kernel in 
linear methods in order
to reach a better estimate of the tachocline width. The basic idea is 
to approximate Eq.~\ref{eq:convol} by a convolution equation \citep{charbonneau99,corbard98}. 
Then, if the 
tachocline profile after inversion is approximated by an $erf$ function 
of width $w$, the `true width' $w_c$ can be obtained by:
\begin{equation}\label{eq:deconvolve}
w_c=\sqrt{w^2-\Delta_r^2}
\end{equation}
where $\Delta_r$ is the radial resolution at  the center of the 
tachocline as defined in Sec.~\ref{sec:inv}.

{\bf (iii)} Non linear or adaptive regularization. 
This approach has been suggested by \citet{corbard98} and fully developed 
by \citet{corbard99}. In brief, 
we construct an iterative process which will adjust 
locally the smoothness term as a function of the gradient 
amplitude found at the previous step. This allow us to keep 
the well constrained sharp gradient zones while still regularizing
elsewhere.

\section{RESULTS FOR THE
EQUATORIAL PLANE AND FOR SPHERICAL AVERAGE}\label{sec:eq}

 All modes have amplitude 
near the equator and therefore we expect to have more chance
to be able to infer tachocline parameters accurately there. 
The first results on the tachocline parameters
were obtained (Tab.~\ref{tab:res}) either at the equator or on
an  average over all latitudes. The equatorial results leads
to an interval $[0.674,0.708]$ or, taking into account uncertainties on both
position and with, $[0.665,0.717]$ for the location of the tachocline.
In the case of  the latitudinally averaged results, we obtain respectively
 $[0.677,0.730]$ and  $[0.671,0.736]$. These first results show that
the equatorial tachocline is centered significantly ($\sim0.02R_\odot$)
below the base of the CZ  and that it is very likely that 
all the strong radial gradient of angular velocity is located below $r_{cz}$
in the equatorial plane. On the other hand, the latitudinally averaged
results suggest that these characteristics are not maintained 
at all latitudes leading, on average, to a thicker tachocline 
centered only around $0.01R_\odot$ below $r_{cz}$. 
Therefore, it seems that, in a latitude-average,
the upper third of the tachocline 
is located in the nearly adiabatically stratified  zone above $r_{cz}$.
Nevertheless this results  must be taken with 
caution because the decreasing radial resolution with latitude  tends
naturally to show a thicker tachocline at high latitudes thereby influencing
the average thickness and position. The parameters obtained this way should 
therefore be regarded as upper limits for the tachocline properties.

The latitude-average width is however probably the best estimate
to use in the formulae obtained from the various theories of 
Sect.~\ref{sec:models} which
assume also a spherical symmetry. In the hydrodynamic
theory, Eq.~\ref{eq:visc} with 
a with of $0.05R_\odot$ leads to an horizontal turbulent viscosity
coefficient of about $3\ 10^6$cm$^2$ s$^{-1}$, several order of magnitude 
higher than the microscopic value. 
In the MHD theories the same value of the width 
leads to a primordial magnetic field strength of 
$|B_0|\sim 10^{-4}$Gauss for both Eq.~\ref{eq:b01} and \ref{eq:b02}.
This suggests that even a weak magnetic field is enough to keep 
the radiative interior rotating rigidly and to confine the radial 
shear  to a thin layer  compatible with observations. 
By analyzing the even splitting coefficients 
from GONG observations (see Sect.~\ref{sec:lowl} below), \citet{basu97}
set an upper
limit of $0.3$MG for a field located at the base of the CZ.
From Eq.~\ref{eq:b01}
this would correspond to a width of $0.0045R_\odot$
which cannot be excluded from observations.
By shearing a poloidal field of $10^{-4}$Gauss the theory
of \citet{mcgregor_charbonneau_99} predicts  the generation of a toroidal
field of $\sim0.1$MG which is also compatible with the  $0.3$MG limit.

\section{VARIATIONS WITH LATITUDE? WITH TIME?}\label{sec:variations}
Theories predict variation with latitude of the shape of the tachocline.
We have seen for example that, in the shallow-water model,
  the presence of a magnetic field would induce a prolate tachocline otherwise
oblate.
Both the magnetic model of  \citet{rudiger} and the 2D hydrodynamic model of 
\citet{spiegel92} predict a thicker tachocline
at the pole than at the equator \citep[see also][]{elliott97}. In the 
model of \citet{gough_mcintyre} the tachocline is thicker where the horizontal
component of the magnetic field in the radiative interior is small. The poles 
should be therefore thicker but that may also be true for other latitudes. 
This means that the shape of the tachocline could also be a diagnostic 
for the geometry of the magnetic field. Some attempts have been made  
to infer latitudinal variations in the tachocline profile from
observations \citep{antia98,charbonneau99}, but no compelling evidence has 
been found yet. Even if, as shown in the previous section, the results 
tend to argue in favor of a prolate tachocline thicker at high latitudes than
at the pole, this may very well be due to the limits in resolution
associated with the datasets used.    

Time variations are also expected mainly because of the changes of
the magnetic field strength and geometry during the solar cycle. Some 
evolution of the large scale circulation inside the layer or the action 
of internal gravity waves may also induce temporal changes or oscillations
in the tachocline parameters. Small amplitude 
oscillations, with a period 
of 1.3 year,  have recently  been  found in the rotation profile 
close to the boundaries of the tachocline \citep{howe2000}. 
These results are based on both GONG and MDI data but have not  
been confirmed by independent analysis \citep{antia_basu2000}. 
If these oscillations are real, the 1.3 year period remains 
unexplained \citep[see][]{gough2000}.
Because of the observed  magnetic cycle,
 we rather expect  the tachocline variations to be on the time scale 
of the solar cycle ($~11$ years). A first attempt
at detecting long term temporal variations in the tachocline
have been made by \citet{basu_schou2000} 
using 11 sets of MDI observations covering 72 days each 
between July 1996 and April 1998. They found
no clear changes in the width or position of the tachocline but 
gave some hints that the position may move slightly outwards 
with increased activity. In the following we use the observations
collected by LOWL instrument during the 6 years of the ascending
phase of the solar cycle (1994-2000) in order to investigate 
the possible cycle related changes in the tachocline. But lets first
introduce this new dataset.

\subsection{Six years of LOWL observations: a new dataset}\label{sec:lowl}
The LOWL instrument \citep{tomczyk95}, based on a Potassium magneto-optical
filter, has been collecting resolved Doppler observations 
for more than six years. 
The spectra were processed using the LOWL pipeline, which has been
recently improved \citep{chano_thesis}. Annual time
series for degrees from $\ell=0$ up to $99$ have been created by using 
spherical harmonic masks  and a
Fast Fourier Transform has been applied to each one of the time series. 
The average duty cycle over one year of observations was
around 20$\%$. 

\begin{figure}
\centering
\includegraphics[width=.8\linewidth]{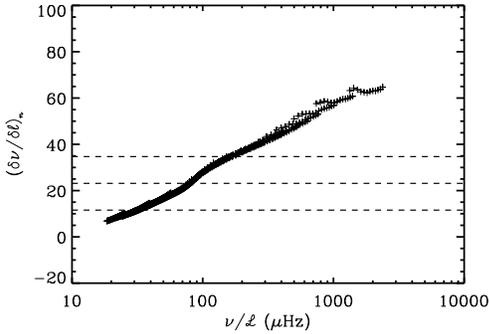}
\caption{Distance in frequency between neighboring $\ell$'s at
constant $n$ as a function of $\nu$/$L$. 
The dashed lines show the frequencies corresponding to
the first three sidelobes. The frequencies used here  are
an average over six independent data sets.\label{fig:dnudl}}
\end{figure}

A detailed description of the fitting method can be found in 
\citet{chano_gong2000_1}. In brief, we have used the general expression of the 
likelihood function, assuming that the statistics of the real and imaginary
parts of the Fourier transform follows a multi-normal distribution, 
described by a covariance matrix 
\citep{schou_thesis,appourchaux_rabello98}. This model is required
due to the fact that the
spherical harmonics 
are not orthogonal over the observed hemisphere. This introduces
a natural leakage between modes with different $\ell$ and $m$.
\begin{figure}[t]
\centering
\includegraphics[width=.8\linewidth,clip=true,bbllx=44bp,bblly=525bp,bburx=355bp,bbury=752bp]{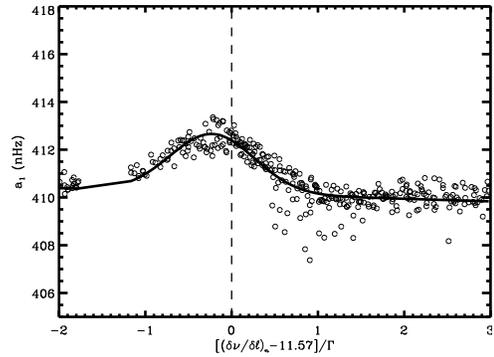}
\caption{Average of $a_1$  over six years.
The solid line denotes the best fit of a Gaussian profile
plus a background given by a straight line.\label{fig:bump}}
\end{figure}

The observed splitting between $m$-components is given in terms of 
$a$-coefficients by:
\begin{equation}
\label{eq:RL}
\nu_{n\ell m} = \nu_{n\ell} + \sum_{j=1}^{n_{coef}}{a_{j}(n,\ell) 
P_{j}^{\ell}(m)} 
\end{equation}
where $P_{j}^{\ell}(m)$ are orthogonal polynomials normalized such that
$P_{j}^{\ell}(\ell)\!\!=\!\!\ell$ \citep[][App.~A]{schou94}. 
The sum over the odd $a$-coefficients gives the rotational splitting
$\delta_{nlm}$ (Eq.~\ref{eq:2D}), while
the even terms are mainly due to magnetic fields and 
second order effects of the rotation. An analysis of the central frequency
$\nu_{nl}$ and the even $a$-coefficients has been carried out using these
data by \citep{chano_gong2000_1}.  These parameters present
significant variations very well correlated with the solar cycle.

One of the main source of systematic errors in the fitting procedure
is the leakage, which affects mainly the $a$-coefficients estimates.
It is particular important when the distance in frequency between 
neighboring $\ell$'s at constant $n$ gets smaller.
Following the 
asymptotic expression given by e.g. \citet{deubner} 
for the frequencies at intermediate and high degrees, we obtain:
$(\frac{\delta\nu}{\delta\ell})_n \simeq \frac{\nu}{2\ell}$.
Therefore, the distance between neighboring $\ell$'s is proportional to
$\nu/\ell$ which, in turn, is related to the 
inner turning point radius $r_t$ through the local sound speed $c(r_t)$ by:
 $2\pi r_t= c(r_t)(L/\nu)$. This property is illustrated on 
Fig.~\ref{fig:dnudl}. Moreover, in the case of observations collected from just 
one site (as LOWL),  sidelobes at 11.57$\mu$Hz from the main peak will  appear 
in the Fourier Transform due to the modulation of one day introduced
in the signal. The leakage between modes
will therefore occur at lower turning point. Using Fig.~\ref{fig:dnudl},
we can predict where to expect complications in the fitting. 
The dashed lines denote the position in
frequency of the first three sidelobes. 
When $\nu/L\approx 31 \mu$Hz the first sidelobe from $\ell\pm1$ 
will cross the position in frequency of the target mode with degree $\ell$.
 The
amplitude of the second 
and  third sidelobes are very small  will not be considered. From the equation above and a standard sound 
speed profile, we can deduce
that  $\nu/L\approx 31 \mu$Hz
corresponds to $r_t=0.9R_\odot$ which is effectively the depth
where the previous analysis of LOWL
data showed an artificial increase of the solar rotation 
\citep[e.g.][]{corbard97,chaplin99}. 
Nevertheless, two different peaks 
very close in frequency can be overlapping or not depending on
their linewidth $\Gamma$. Therefore, the  $a$-coefficients should be plotted
against $[(\delta\nu/\delta\ell)_n -11.57]\Gamma^{-1}$
(Fig.~\ref{fig:bump}).
An important feature or $bump$ appear centered approximately at zero.
Its amplitude is  maximum  
for $a_1$ ($\sim 2$nHz). The same analysis was carried out 
year by year, showing that this feature
is found similar for all time series. Thus, we decided to remove 
this systematic 
error by fitting the $bump$ with a Gaussian profile in 
 the average over six years and by  subtracting afterwards
 this profile from every yearly fit. 
 The even $a$-coefficients do not
show similar $bumps$ and have therefore not been changed. 




\subsection{Results and Discussion}

\begin{figure}[t]
\hskip -1cm
\includegraphics[angle=90,height=.9\linewidth]{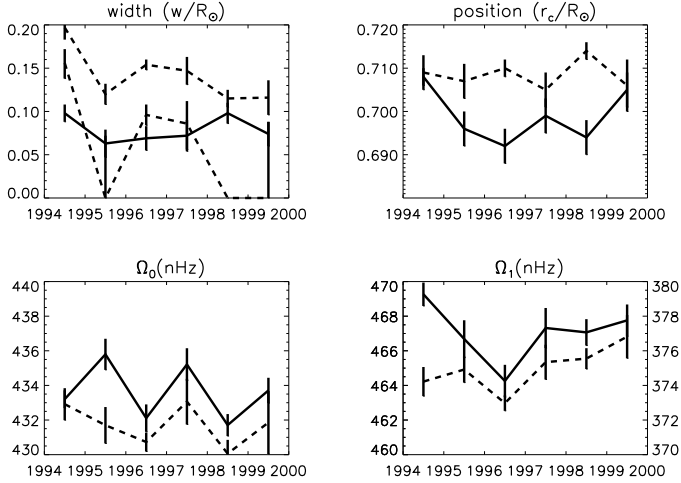}
\caption{Inferred tachocline parameters as a function of time from LOWL observations. The full and dashed lines correspond respectively
to fit at the equator and at $60^\circ$. For the width (upper left), the lower
dashed line is obtained after deconvolution using Eq.~\ref{eq:deconvolve}. The scale on the right of the lower right panel corresponds to the fit at $60^\circ$ (dashed line).
\label{fig:param}}
\end{figure}

\begin{figure}[t]
\hskip -1cm
\includegraphics[angle=90,height=.35\linewidth,width=1.2\linewidth]{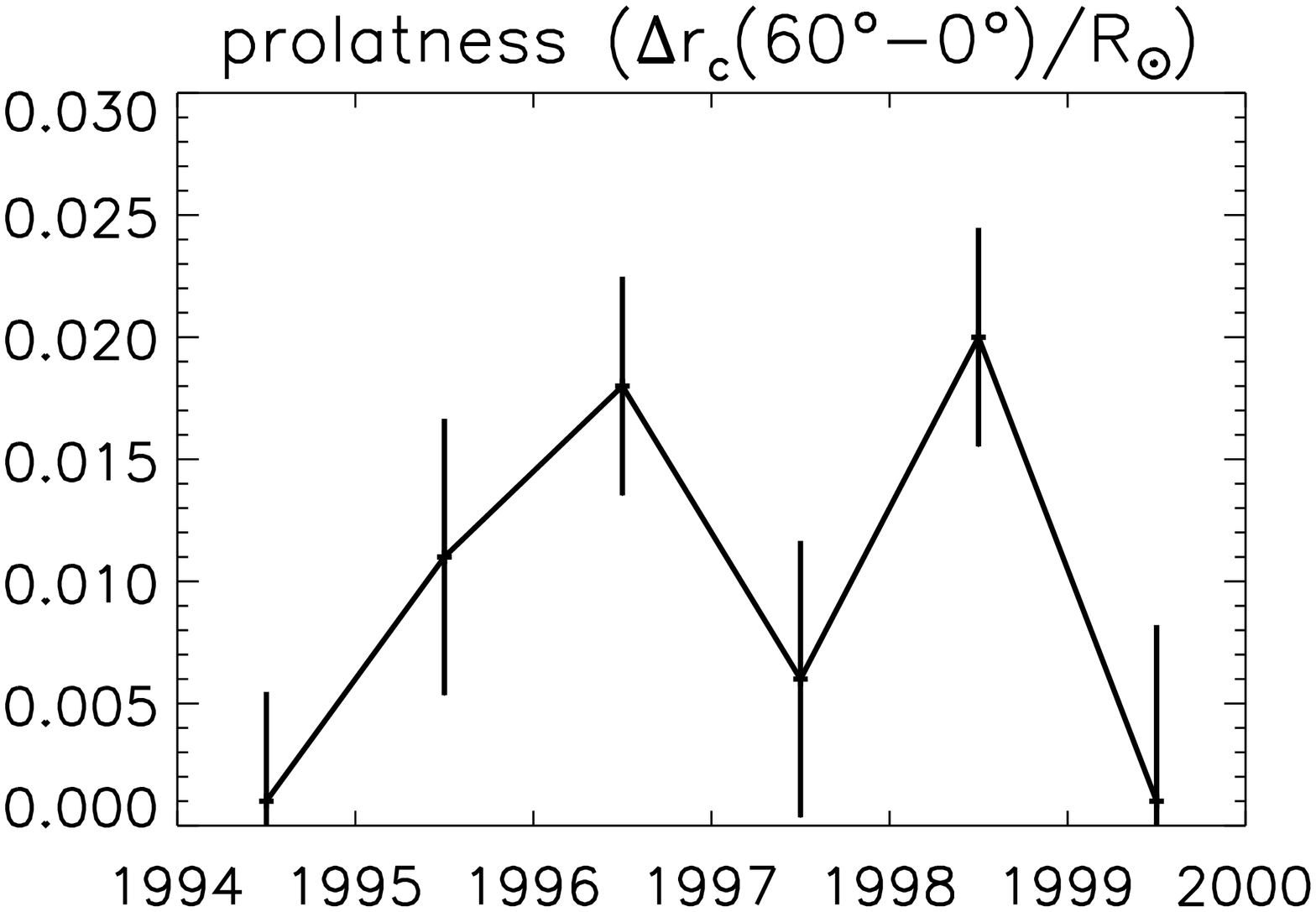}

\vskip -.8cm

\hskip -1cm
\includegraphics[angle=90,height=.35\linewidth,width=1.2\linewidth]{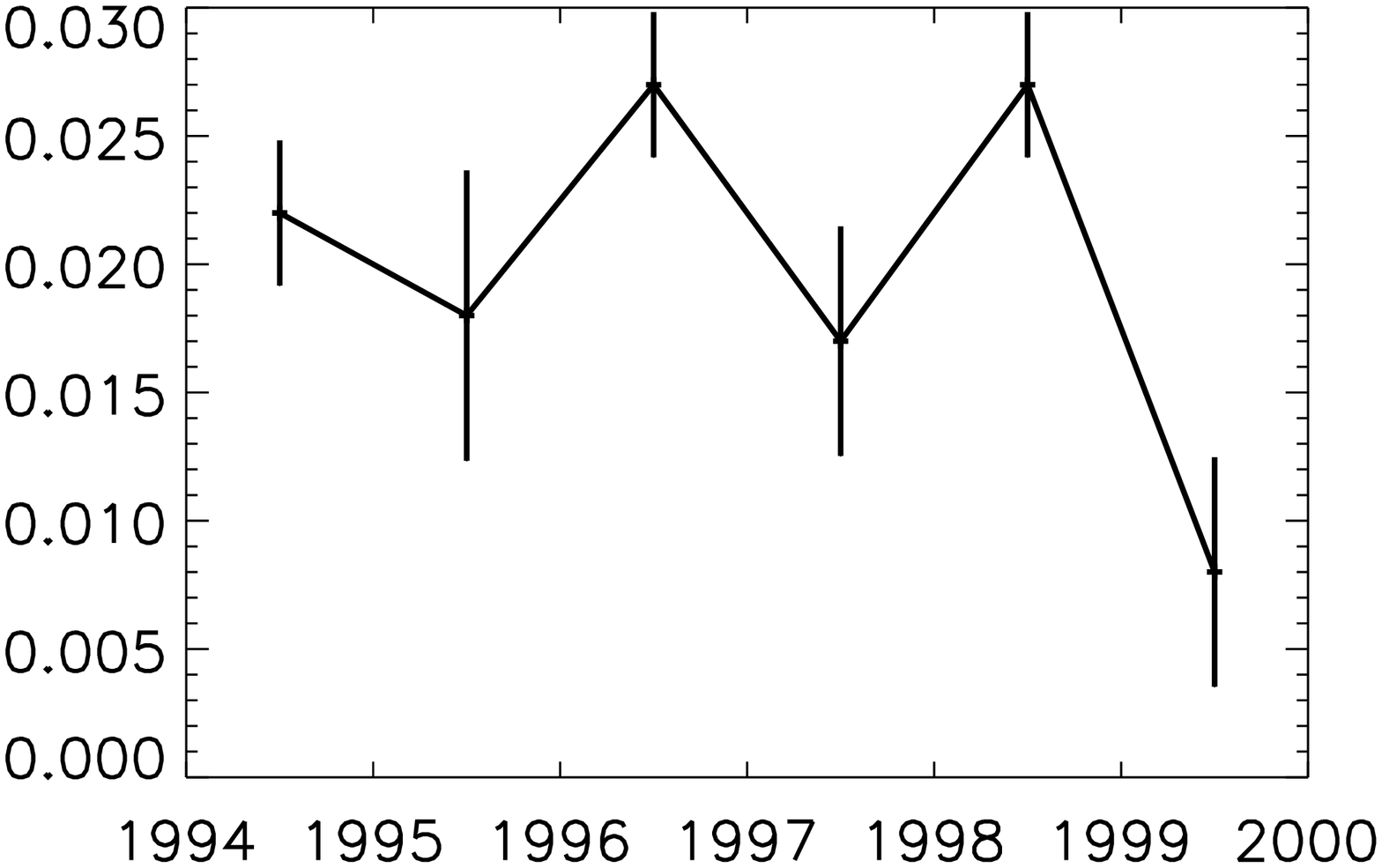}
\caption{ Difference
between the central position  infered at  $60^\circ$ and at 
the equator, as a function of time.  The upper plot corresponds to 
the results of Fig.~\ref{fig:param} while the lower plot corresponds 
to inversion using stronger regularization. \label{fig:prolate}}
\end{figure}

The splitting coefficients obtained for each year have been inverted 
using a 2D RLS inversion code \citep{corbard97}. The rotation profile 
has then been fitted at the equator and $60^\circ$ by an $erf$ function 
as given by Eq.~\ref{eq:params}. The averaging kernels at the center
of the tachocline have been
fitted by a gaussian in order to estimate  the radial resolution.
The results are shown in Fig.~\ref{fig:param}. The radial 
resolution obtained is about $0.12R_\odot$ at $60^\circ$ and
$0.10R_\odot$ at the equator. The angular resolution achieved
is related to the number of $a$-coefficients in Eq.~\ref{eq:RL}.
Our analysis includes up to 9 coefficients leading to a
maximum angular resolution of about $20^\circ$ at the equator.
Because the radial shear doesn't exist at about $30^\circ$ and
because the angular resolution decrease at high latitudes, we limit
our analysis to two latitudes: the equator and $60^\circ$. Therefore,
in the following, prolateness refers to the difference 
 between the central position of the layer at these two latitudes.

 Because the inferred width at the equator
is always lower than the resolution we cannot use our simple
model for deconvolving. This  indicates that the width of the tachocline
at the  equator is probably lower than the local spacing of the
 grid used for the inversion i.e. $0.02R_\odot$.  The same happens
at $60^\circ$ for 1995, 1998 and 1999. There is therefore no 
strong evidence of a systematically thicker tachocline at $60^\circ$.   
Moreover the errors reported on the plot are formal errors as obtained
from the fits but Monte Carlo simulations have been carried out which suggest 
that the uncertainties on the inferred widths after the whole inversion
process are between $\pm0.02$ and $\pm0.03R_\odot$. 

The center 
of the tachocline is always found deeper at the equator than at $60^\circ$.
The variation of this prolateness with time is shown in
Fig.~\ref{fig:prolate}. The maximum of prolateness found is 
about $0.02R_\odot$ and
no prolateness is found the first and last years. These differences
 and their fluctuations are nevertheless very small and are also very sensitive
to the inversion parameters chosen and especially the amount 
of regularization used. The lower panel of Fig.~\ref{fig:prolate}
illustrates this point by showing that, with a more regularized inversion,
the maximum prolateness is about $0.03R_\odot$ and a minimum is no longer
found for 1994.

Generally speaking, we do not find from this analysis evidence of any general
trend or significant oscillation in the tachocline parameters during
the ascending phase of the actual solar cycle. In particular we do not find
an outward trend for the central position as suggested 
by \citet{basu_schou2000}. Nevertheless, a
 prolateness of the 
layer is observed every year which allow us to,  at least,  exclude
an oblate tachocline. The amount of prolateness found between the equator
and $60^\circ$ is however of the same order of magnitude than the 
uncertainties on the width of the layer. We can therefore only set
 an upper limit for the prolateness which is around $0.03R_\odot$.
This is in good agreement with previous estimates of \citet{antia98}
($0.004\!\!<\!\!\Delta_{rc}/R_\odot\!\!<\!\!0.023$) and \citet{charbonneau99} 
($\Delta_{rc}\simeq 0.023R_\odot$). Using the shallow-water model
and shape curves as shown in Fig.~\ref{fig:prolat_th},
this would correspond to a toroidal magnetic field strength
of about $0.1$MG if it is located in the overshoot layer or 
about $0.6$MG if it is located in the radiative interior.
If the toroidal field is concentrated in bands migrating towards 
the equator during the ascending phase of the cycle, one would expect, from
the shallow water model, a decreasing prolateness for the period of LOWL
observations. 
This general trend cannot be excluded
but is not  observed from our preliminary analysis of LOWL data.


\section*{Acknowledgments}
T. Corbard and S.J. Jim\'enez-Reyes are very thankful to the organizers
of the meeting for providing financial support.  T. Corbard acknowledge support from NASA grant S-92678-F.


%
%

\end{document}